# WHAT IS THE SMALL–SCALE VELOCITY DISPERSION OF GALAXY PAIRS?

LUIGI GUZZO
*Osservatorio di Brera, Via Bianchi 46, I-22055, Merate (CO), Italy*

KARL B. FISHER & MICHAEL A. STRAUSS
*Institute for Advanced Study, School of Natural Sciences, Princeton, NJ 08540, USA*

RICCARDO GIOVANELLI & MARTHA P. HAYNES
*Department of Astronomy, Space Sciences Building, Cornell University, Ithaca, NY 14853, USA*

ABSTRACT

We present some results from an ongoing investigation of the anisotropy of the two–point correlation function for optical galaxies. We have estimated $\xi(r_p, \pi)$ from the Perseus-Pisces redshift survey, which is now virtually 100% complete for all morphological types to $m_B = 15.5$. We detect strong distortions of the iso-correlation contours in $\xi(r_p, \pi)$. These correspond to a global ('cluster' plus 'field') pairwise velocity dispersion at 1 $h^{-1}$ Mpc, $\sigma(1)$, which is about a factor of two higher than the canonical value $\sim 350$ km s$^{-1}$ obtained from the CfA1 survey and recently from the 1.2 Jy IRAS redshift survey. However, the variance of this value is still rather strong even within the relatively large volumes surveyed (typically $90° \times 45° \times 100$ $h^{-1}$ Mpc). From the largest volume explorable (126 $h^{-1}$ Mpc depth), the best bet for the 'cosmic' value of $\sigma(1)$ seems to be between 600 and 700 km s$^{-1}$. However, for a hopefully robust determination of its value we will have to wait for the analyses of the next generation of larger **optical** redshift surveys (such as, e.g., the now completed ESP[1] or the forthcoming Sloan Digital Sky Survey[2]).

## 1. Introduction

Peculiar velocities distort the maps of the galaxy distribution when radial velocities are used directly as a measure of their distance through the Hubble relation. The observed distortions contain a wealth of important information on the statistical properties of large–scale motions, hence on the true mass distribution. In particular, the two–point correlation function in redshift space $\xi(s)$ differs from that in real space $\xi(r)$ in two respects: on small scales correlations are suppressed due to the virialized motions in rich clusters, which in redshift space elongate structures along the line of sight; on large scales coherent motions produced by infall into overdense regions or by outflow out of underdense regions have the opposite effect, i.e. they enhance correlations. The standard way to quantify these effects is to estimate correlations as a function of two variables, namely the two components of the separation vector on the plane of the sky and along the line of sight, indicated as $\xi(r_p, \pi)$. Projecting $\xi(r_p, \pi)$ onto the $r_p$ axis allows us to reconstruct the undistorted real space correlation function. In addition, a proper modeling of the observed $\xi(r_p, \pi)$ allows us to characterize the pairwise velocity distribution function along the line of sight in terms of its moments. The first and second moments play a major role in describing the observed distortions on $\xi(r_p, \pi)$, being related, respectively, to the large-scale compression [the mean streaming $v_{12}(r)$] and to the small-scale stretching [the dispersion $\sigma(r)$]. In general, both moments are a function of the separation $r$ of the pairs. Here we shall report on new results concerning the second moment $\sigma(r)$.

The standard quoted value for $\sigma(r)$ at small separations [in particular at 1 $h^{-1}$ Mpc, $\sigma(1)$ ] has long been that estimated by Davis & Peebles[3] on the CfA1 survey, $\sigma(1) = 340 \pm 40$ km s$^{-1}$. The recent determination of $\sigma(1)$ from the IRAS 1.2 Jy redshift survey[4] seemed to provide a confirmation of such a value, with $\sigma(1) = 317^{+40}_{-49}$ km s$^{-1}$. However, recent re-analyses of the CfA1 survey[5,6] showed that the original value was rather sensitive to local corrections, like that for Virgocentric infall, pointing to a value about twice as high. As we shall show here, the analysis of an optically selected survey with a volume significantly larger than that of CfA1 confirms that the value of $\sigma(1)$ is most probably higher than the original estimate. The uncertainty of $\sigma(r)$ due to variations in the sample size is however still quite strong, clearly indicating the need of larger surveys for obtaining a more robust estimate of its average 'cosmic' value.

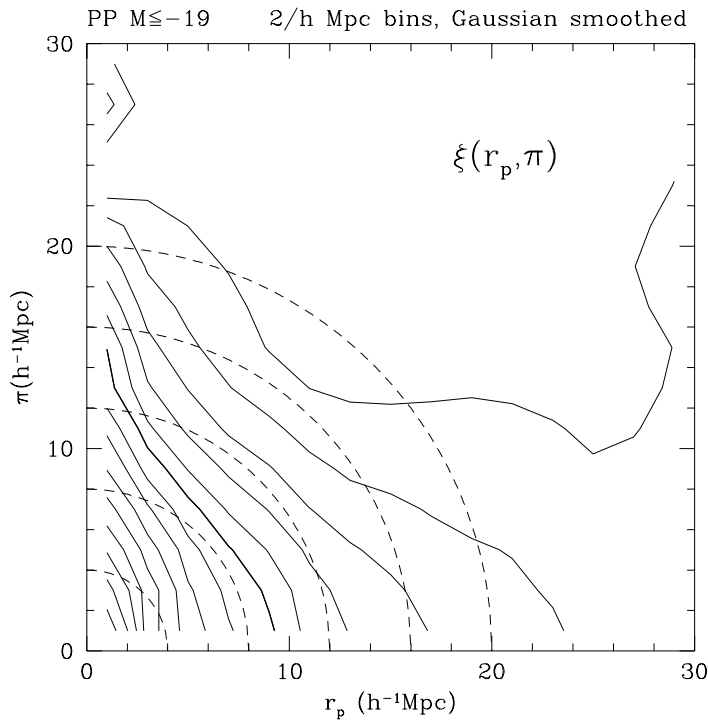

Fig. 1. $\xi(r_p, \pi)$ for the sample with M $\leq$ -19 and $d \leq 79$ $h^{-1}$. The heavy contour corresponds to $\xi = 1$. For larger values of $\xi$, the contours are logarithmically spaced, with $\Delta \log \xi = 0.1$. Below, they are linearly spaced with $\Delta \xi = 0.2$ down to $\xi = 0$.

## 2. The Observed $\xi(r_p, \pi)$

We defined $r_p$ and $\pi$ as in[4], and used the standard estimator[3] to obtain $\xi(r_p, \pi)$ for

volume–limited subsamples of the Perseus–Pisces (PP) redshift survey. (For a detailed description of the PP redshift catalogue, see e.g.[7]). We stress the importance of using strictly volume–limited subsamples for surveys like PP (but also CfA2), where a dominant perpendicular structure is located near the peak of the selection function. In our case, this has the beneficial effect of suppressing the dominance of the Perseus–Pisces chain. It is interesting to note that in this case both the two–point correlations[8] and the higher moments [7,9] agree between PP and CfA2. However, as we shall see, our results on velocity distortions indicate that these samples are still too small for providing the correct 'cosmic' mixture of cluster and non–cluster galaxies, necessary to estimate a global average value for $\sigma(r)$ at small separations (similar to what can be obtained from large N–body simulations).

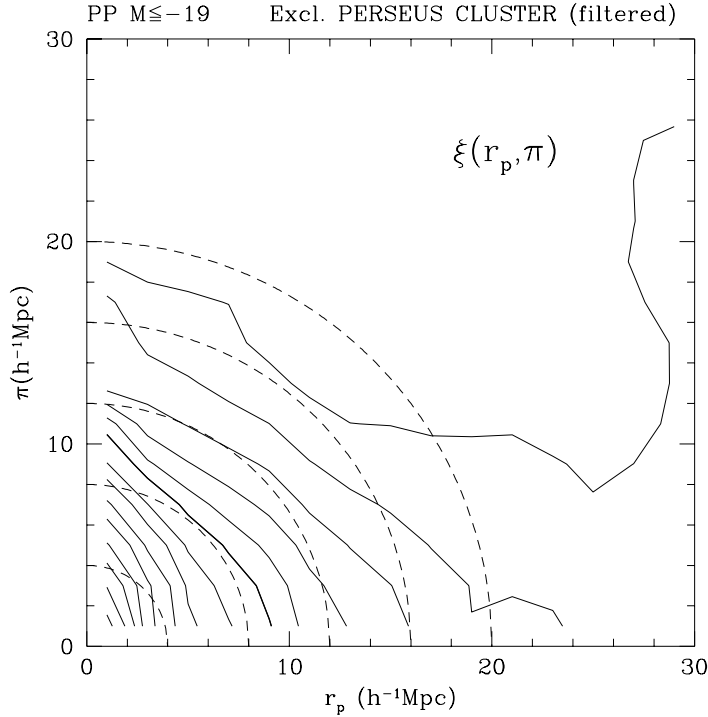

Fig. 2. Same as figure 1, but limiting the sample to RA $< 3^h$, i.e. excluding the Perseus cluster.

Figure 1 shows $\xi(r_p, \pi)$ for a sample limited to $M \leq -19$ (using $H_o = 100\,\mathrm{km\,s^{-1}\,Mpc^{-1}}$) and $d \leq 79\,h^{-1}\,\mathrm{Mpc}$, extending from $22^h$ to $4^h$ in RA and from $0°$ to $45°$ in DEC, containing 1159 galaxies. $\xi(r_p, \pi)$ has been estimated using a $2\,h^{-1}\,\mathrm{Mpc}$ binning, and for clarity the displayed maps have been smoothed with a 2–D Gaussian filter. Note the strong elongation of the contours along the line of sight at small $r_p$'s, while on larger scales the opposite effect is evident. Part of this large–scale compression is most probably produced by streaming towards the PP supercluster. However, part of the effect is probably also due to the true spatial distribution of the pairs within the PP supercluster, not adequately compensated by other structures along different directions. In this situation, the two contributions cannot be disentangled using the redshift space data alone, thus discouraging any serious attempt to estimate the mean streaming on large scales.

Figure 2 shows the effects on $\xi(r_p, \pi)$ of simply restricting the sample to RA $\leq 3^h\ 10^m$, thereby excluding the richest cluster in the survey, the Perseus cluster. Although only 117 galaxies have been eliminated by this selection, the effect on $\xi(r_p, \pi)$ is dramatic, significantly reducing the small–scale distortion. In the next section we shall quantify this change in terms of the implied pairwise velocity dispersion.

## 3. The Small–Scale Pairwise Velocity Dispersion

A proper modelling of the distortions induced on $\xi(r_p, \pi)$ by peculiar velocities has to take into account the effects of, at least, the first two moments of the velocity distribution function $f$. Once a suitable functional dependence of $f$ on the streaming $v_{12}(r)$ and dispersion $\sigma(r)$ is chosen, the problem reduces to finding a physically motivated parametrization of the scale dependence of these two quantities. This problem is discussed in detail in [4]. Here we present results obtained using the Davis & Peebles original streaming model, based on the similarity solution of the BBGKY equations. $\xi(r_p, \pi)$ is modelled as the convolution of the real space correlation function $\xi(r)$ with the distribution function $f$. In the nonlinear regime, an appropriate choice for $f$ is an exponential, so that one can write

$$1 + \xi(r_p, \pi) = H_\circ \int_{-\infty}^{+\infty} dy\, [1 + \xi(r)]\, \frac{1}{\sqrt{2}\sigma(r)} \exp\left\{ -\sqrt{2} H_\circ \left| \frac{\pi - y\left[1 + \frac{v_{12}(r)}{H_\circ r}\right]}{\sigma(r)} \right| \right\}\ ,\quad (1)$$

where $r^2 = r_p^2 + y^2$, while the scale dependence of the streaming is given by the model[3]

$$v_{12}(r) = -H_\circ r \frac{F}{1 + \left(\frac{r}{r_o}\right)^2}\ .\quad (2)$$

The other essential ingredient here is the assumption of a slowly varying $\sigma(r)$, so that the second moment can be treated as a single free parameter. Given the observed $\xi(r_p, \pi)$, we first estimate the real–space correlation function by fitting a power–law model to the projected function $w_p(r_p)$ [4]. Then we fit the above model to slices through the observed $\xi(r_p, \pi)$ at fixed $r_p$, fitting simultaneously for the amplitude of the streaming $F$ and the dispersion $\sigma$. All the fits are performed using the maximum likelihood method discussed in[10], based on 400 bootstrap resamplings. We limit the fit to small values of $r_p$, where the velocity dispersion dominates over the mean streaming and the above model can be considered a reasonable approximation to reality.

In Figure 3 we show the results of fitting the model to the first bin of the $\xi(r_p, \pi)$ of Figure 1, i.e. the slice with $0 \leq r_p \leq 2\ h^{-1}$ Mpc. The best fit values correspond to $F = 0.54^{+1.77}_{-1.87}$ and to a dispersion $\sigma = 769^{+171}_{-342}$ km s$^{-1}$. The fit is limited to $\pi \leq 10$ h$^{-1}$ Mpc, i.e. where the power–law model for $\xi(r)$ is a good representation of the data. However, the value of $\sigma$ is robust to changing the upper fitting limit to 20 or 30 $h^{-1}$ Mpc, with variations of a few tens of km s$^{-1}$.

However, using the $\xi(r_p, \pi)$ map of Figure 2, i.e. excluding the Perseus cluster, the estimated $\sigma$ drops by nearly 200 km s$^{-1}$, as shown in Table 1. In the same table we list also the results obtained on two deeper sub-samples, up to 126 $h^{-1}$ Mpc, again over the whole RA range. The aim here is to look for convergence of $\sigma(1)$ as the volume increases. As can be seen, the estimated values still show strong fluctuations at the largest explorable sizes.

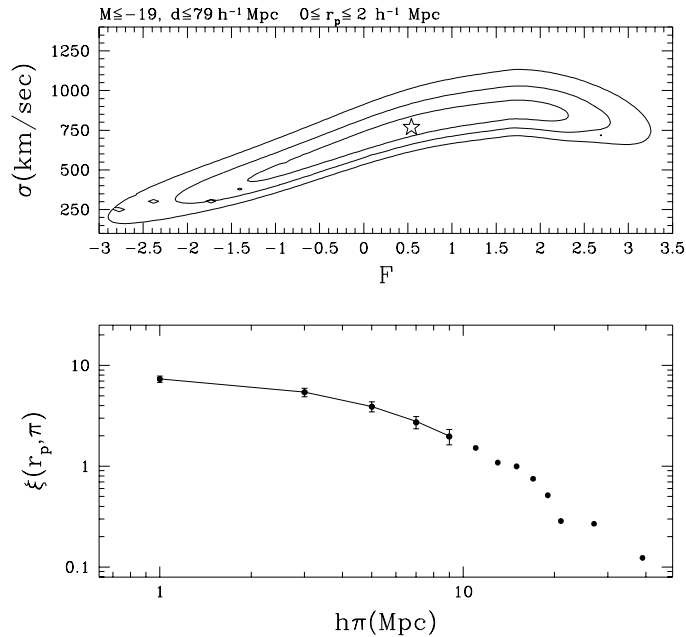

Fig. 3. Fit of the model $\xi(r_p, \pi)$ to the data from the $M \leq -19$ sample. Upper panel: Maximum Likelihood contours corresponding to the 68.3%, 95.4%, and 99.7% confidence levels on the two parameters taken separately. Lower panel: data points from the slice of the observed $\xi(r_p, \pi)$, with the fitting model superimposed.

Also, the lower 68% confidence limit is rather large for all the estimates but that from the $M \leq -19$ "No Perseus" sample. Globally, one could dare to say that there is a tendency towards a value $\sim 700$ km s$^{-1}$, but this is certainly not statistically significant from these data. We are working towards improving the description of the streaming. However, as discussed by various authors[4], the value of the second moment is rather robust with respect to variations in the streaming model, and we do not expect the best-fit values of $\sigma(1)$ to change significantly. A full description of the work outlined here will be presented in a forthcoming paper.

LG thanks J. Bahcall for hospitality at the IAS, where this work was initiated.

| Sample | | $N_{GAL}$ | $\sigma(1)$ (km sec$^{-1}$) |
|---|---|---|---|
| $M \leq -19.0$ | $hd \leq 79$ Mpc | 1159 | $769^{+171}_{-342}$ |
| Same, no Perseus | | 1042 | $613^{+73}_{-57}$ |
| $M \leq -19.5$ | $hd \leq 100$ Mpc | 943 | $927^{+167}_{-539}$ |
| $M \leq -20.0$ | $hd \leq 126$ Mpc | 612 | $717^{+190}_{-427}$ |

Table 1. ML values for the pairwise velocity dispersion at separations between 0 and 2 h$^{-1}$ Mpc

## 4. References

1. G. Vettolani et al., in *Wide–Field Spectroscopy and the Distant Universe*, S.J. Maddox & A. Aragón–Salamanca eds., Singapore: World Scientific (1995), in press
2. J. Gunn, & D.H. Weinberg, in *Wide–Field Spectroscopy and the Distant Universe*, S.J. Maddox & A. Aragón–Salamanca eds., Singapore: World Scientific (1995), in press
3. M. Davis & P.J.E. Peebles, *ApJ* **267** (1983) 465.
4. K.B. Fisher, M. Davis, M.A. Strauss, A. Yahil & J.P. Huchra, *MNRAS* **267** (1994) 927.
5. H.J. Mo, Y.P. Jing & G. Börner, *MNRAS* **264** (1993) 825.
6. W.H. Zurek, M.S. Warren, P.J. Quinn & J.K. Salmon, in *Cosmic Velocity Fields*, F.R. Bouchet & M. Lachièze–Rey, Paris, Frontiéres (1993), p. 465.
7. S. Ghigna, S. Borgani, S.A. Bonometto, L. Guzzo, A. Klypin, J.R. Primack, R. Giovanelli,, & M.P. Haynes, *ApJ* **437** (1994) L71.
8. E. Branchini, L. Guzzo & R. Valdarnini, *ApJL* **424** (1994) L5.
9. M.S. Vogeley, M.J. Geller, C. Park & J.P. Huchra, *ApJ*, in press.
10. K.B. Fisher, M. Davis, M.A. Strauss, A. Yahil & J.P. Huchra, *MNRAS* **266** (1994) 50.